\documentclass[showpacs,prl,twocolumn,superscriptaddress]{revtex4}
\usepackage{graphicx}
\newcommand{\etal}{{\it et al.}}

\begin{document}

\title{Linear dichroism and the nature of charge order in underdoped cuprates}

\author{M. R. Norman}
\affiliation{Materials Science Division, Argonne National Laboratory, Argonne, IL 60439, USA}

\begin{abstract}
Recent experiments have addressed the nature of the charge order seen in
underdoped cuprates.  Here, I show that x-ray absorption and linear dichroism are an excellent
probe of such order.  Ab-initio calculations reveal that a d-wave charge density wave order
involving the oxygen ions is a much better description of the data than alternate models.
\end{abstract}

\date{\today}

\maketitle
The observation of charge order in underdoped YBCO by NMR \cite{wu} followed by its
observation by both soft \cite{keimer} and hard \cite{chang} x-ray scattering, along with
the probable role of the charge order in Fermi surface reconstruction \cite{sebastian},
has been one of the
most significant developments in the field of high temperature cuprate superconductivity in
the past few years.  Of particular note was the suggestion that the charge order was
of a novel d-wave nature, involving charge modulations that are out of phase on
the two oxygen ions in a given CuO$_2$ unit \cite{sachdev}.  Such a d-wave phasing
relation has been demonstrated by Fourier transformed STM data \cite{davis} and is
consistent as well with
the azimuthal dependence of resonant x-ray scattering at the Cu L$_3$
edge \cite{comin}.  The latter involves modeling the x-ray data as due to a spatial
modulation of the energies of the copper 3d orbitals relative to the oxygen
2p ones \cite{ackhar}.

Related to these measurements was a claim of seeing x-ray natural circular dichroism
(XNCD) at the Cu K edge in underdoped Bi2212 \cite{kubota}.  An XNCD signal has been
claimed as well in underdoped LBCO \cite{he}.  Although chiral charge order could cause an
XNCD signal \cite{mike13}, a more natural explanation of the data is linear dichroism
contamination, which is always present in circularly polarized beams \cite{sergio07}.  This has
been confirmed by recent measurements at the Cu K edge in Bi2212, where
the expected azimuthal angular dependence of $\cos(2\psi)$ for linear dichroism was seen in 
the ``XNCD" signal \cite{lee}.

Here, I show that in fact linear dichroism itself provides an exquisite probe of charge order
in the cuprates if the ordering is one-dimensional in nature (stripes) as opposed to
two-dimensional in nature (checkerboards) - in the latter case, there is no linear dichroism
due to charge order.  For calculational purposes, I consider Hg1201 since that material is
tetragonal, meaning there is no structural linear dichroism to contend with.  Calculations
were performed with the multiple scattering Greens function code FDMNES \cite{joly}
including spin-orbit interactions \cite{so}, for various cluster radii.
Atom positions for Hg1201 were taken from Wagner \etal~\cite{wagner}.
Results shown here are for the Cu L$_2$ and L$_3$ edges for a cluster radius
of 5 $\AA$ (36 atoms surrounding each absorbing copper site), with calculations also
performed at the Cu K and Cu L$_1$ edges.

\begin{figure}
\includegraphics[width=0.75\hsize]{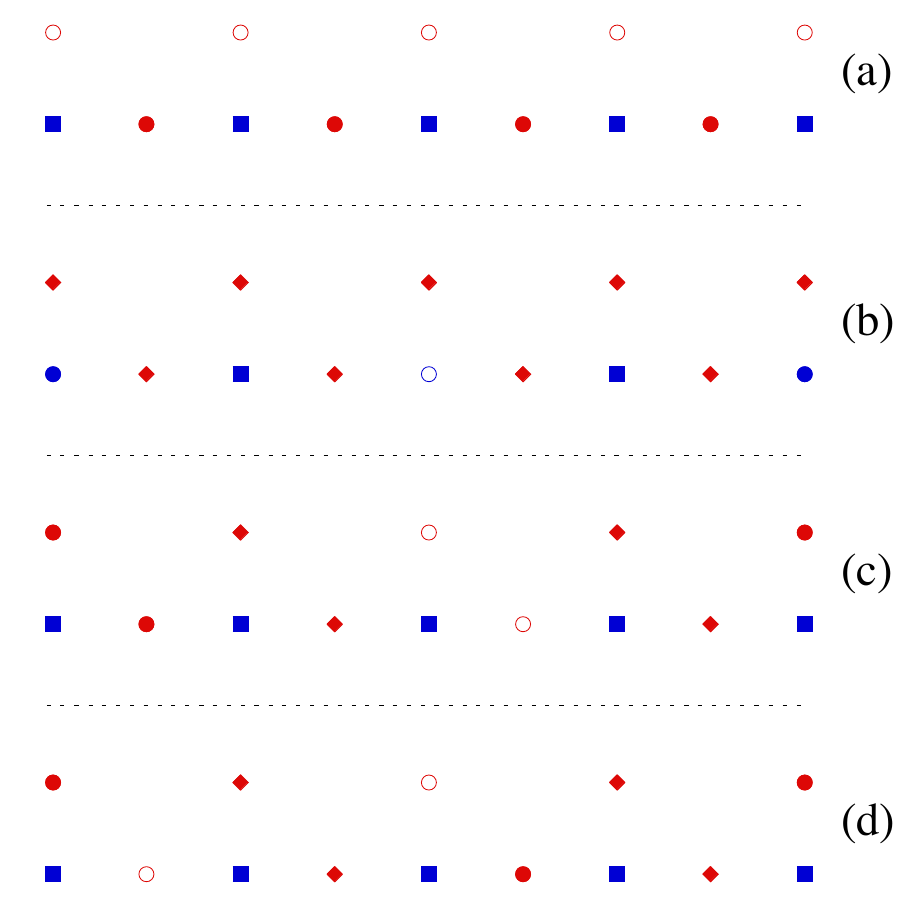}
\caption{(Color online) Charge modulation patterns assumed in this work
for a charge ordering vector of 0.25 reciprocal lattice units.  (a) is for nematic order,
with (blue) squares the unmodulated copper sites, and filled and open (red) circles
oxygen sites with an excess (deficit) of 2p electrons. (b) is for a copper charge
modulation, with (blue) squares the unmodulated copper sites, and filled and open (blue)
circles copper sites with an excess (deficit) of 3d electrons, with (red) diamonds
the unmodulated oxygen sites.  (c) is for an oxygen s-wave charge modulation, with 
(blue) squares the unmodulated copper sites, and filled and open (red) circles oxygen
sites with an excess (deficit) of 2p electrons, with unmodulated oxygen sites as
(red) diamonds.  (d) is for an oxygen d-wave charge modulation, with the same
notation as in (c).
}
\label{fig1}
\end{figure}

Fig.~1, I show the four ordering patterns assumed: (a) nematic order within a CuO$_2$
unit (the two oxygen ions being out of phase), (b) charge modulation on the copper sites, 
(c) s-wave charge modulation on the oxygen sites (the two oxygen ions in each CuO$_2$
unit are in phase) and (d) d-wave charge modulation on the oxygen sites (the two oxygen 
ions in each CuO$_2$ unit are out of phase).  To simplify the calculations, a charge
ordering wavevector of 0.25 in reciprocal lattice units was assumed.
The results presented involve a convolution of the calculated spectrum with 
both a core hole (1.66 eV) and a photoelectron
lifetime, with the latter having a strong energy dependence (at high energies, 15 eV, with a
midpoint value at 30 eV above the Fermi energy).

\begin{figure}
\includegraphics[width=\hsize]{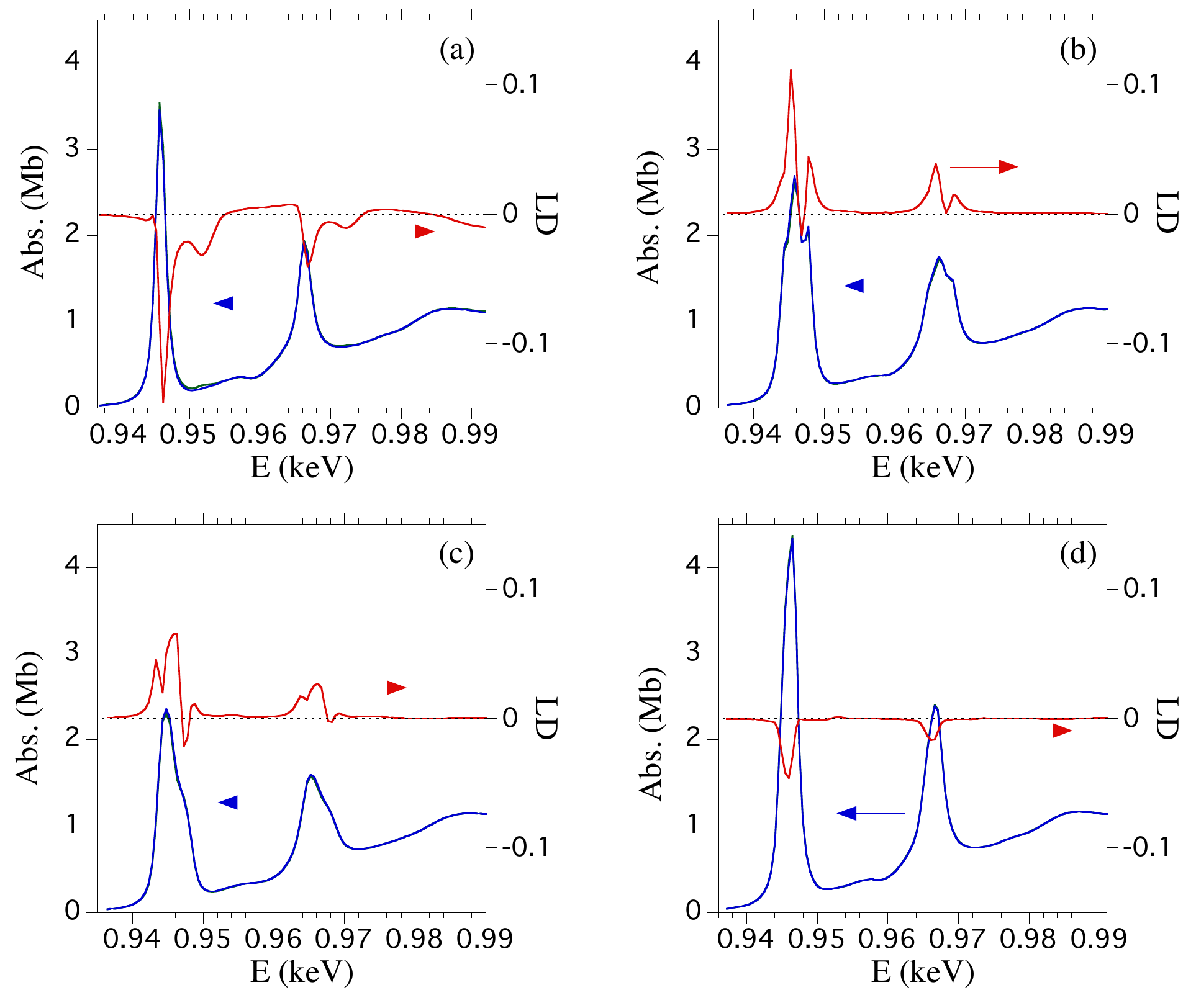}
\caption{(Color online) X-ray absorption at the Cu L$_3$ and L$_2$ edges for a $k$
vector along the c-axis for Hg1201.  Two absorption curves are shown, one for the electric field
vector along x, the other along y.  The linear dichroism (LD) is taken as the difference
for the electric field along y and that along x, with the charge modulation wavevector
along x.  Each panel (a)-(d) corresponds to the modulations shown in Fig.~1, with
a maximum modulation of $\pm$ 0.05 electrons.
}
\label{fig2}
\end{figure}

In Fig.~2, results for the Cu L$_3$ and L$_2$ edges are shown for the four cases
illustrated in Fig.~1 assuming a charge modulation of $\pm$0.05 electrons as assumed in
earlier modeling \cite{ackhar}.  Here, the x-ray $k$ vector is directed
along the c axis \cite{shift}.
The absorption for nematic order (a) is similar to that without any
charge modulation, except for the presence of a substantial dichroism near each
absorption peak (almost 5\% of the absorption maximum at the L$_3$ edge).  For
copper charge modulation (b) one finds substantial broadening of the two absorption
peaks due to splitting of the peaks from the fact that there are now three different copper
sites (Fig.~1b).  Similar broadening is also seen for the s-wave oxygen modulation
case (c).  For the d-wave modulation case (d), though, one finds absorption peaks whose widths are
only moderately larger than without charge order, with a significantly reduced linear
dichroism (1\% of the absorption maximum at the L$_3$ edge).

\begin{figure}
\includegraphics[width=\hsize]{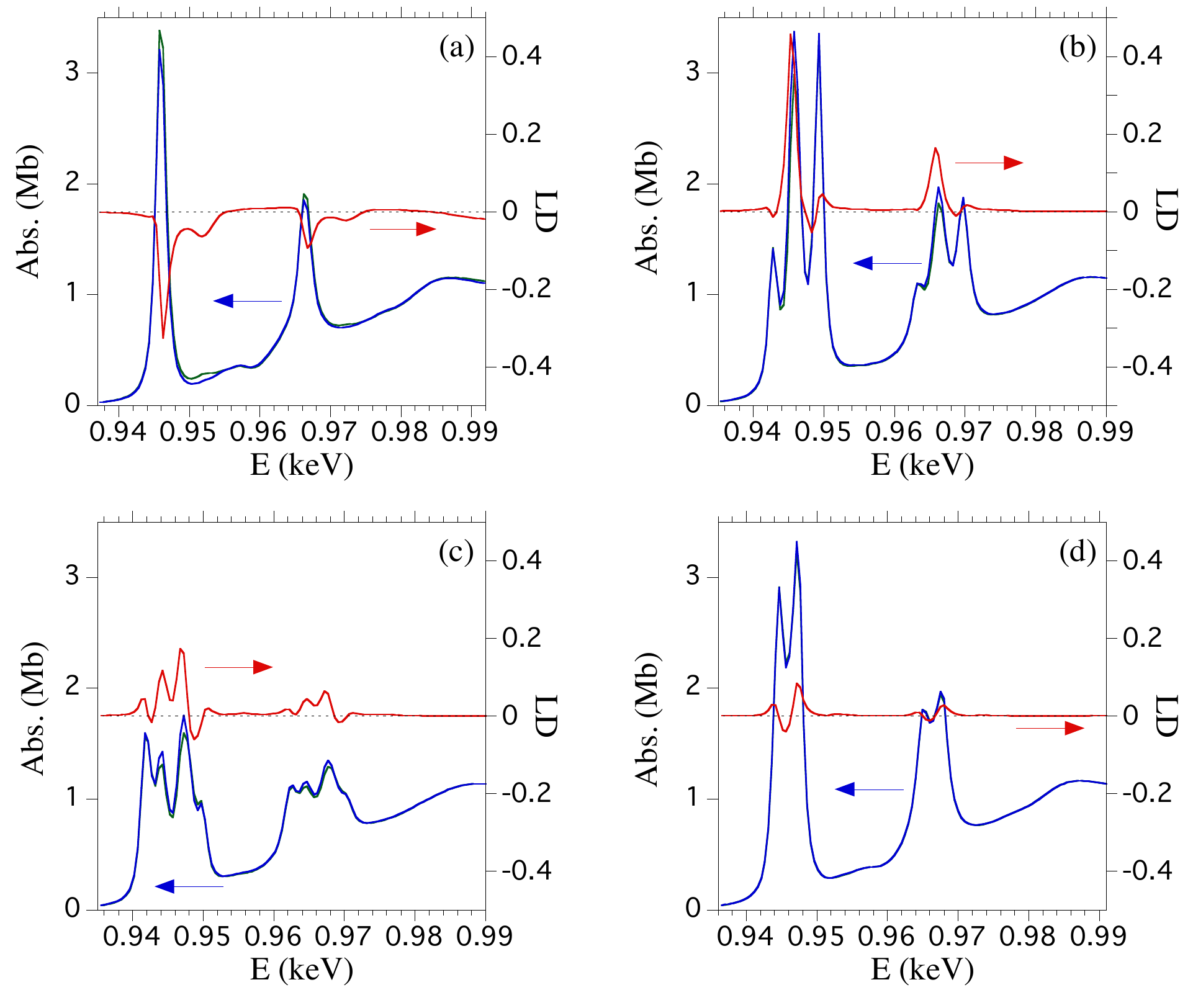}
\caption{(Color online) Same as Fig.~2, but for a maximum charge modulation of $\pm$ 0.1 electrons.
}
\label{fig3}
\end{figure}

In Fig.~3, results are shown analogous to Fig.~2, but for a charge modulation of $\pm$0.1 electrons instead.
For this increased value, one sees substantial splitting of the absorption peaks in all cases except
for the case of nematic order.  In addition, the linear dichroism signal is enhanced by roughly a
factor of four.  Although we do not show results for the Cu K and L$_1$ edges, in those cases
as well, a substantial dichroism exists.  The shape of the absorption peaks is also altered relative to that
without charge order, and this
is most pronounced for the case of an s-wave oxygen modulation.

The observed absorption peaks for the L$_3$ edge of underdoped cuprates are actually quite narrow.
For the case of Hg1201, the FWHM is around 1.2 eV with no sign of splitting of the peak \cite{tabis}.
If anything, even narrower peaks are seen in underdoped YBCO \cite{keimer} and Bi2201 \cite{comin2}.
The simulation results presented here
are actually broader than what is observed, even in the case of no charge modulation, and this could be
corrected by reducing
the assumed core hole broadening.  Regardless, the narrowness of the peaks and the absence of any
splitting is definitely inconsistent with either a charge modulation on the copper sites, or an s-wave
modulation on the oxygen sites.  As such, the results presented here are completely consistent with a d-wave pattern
as advocated in previous work \cite{davis,comin}.  Further information could be obtained if the light spot
imaged a single charge ordering domain, since for each charge modulation pattern, the linear dichroism 
has a unique energy profile.

These results can easily be extended by looking at resonant x-ray scattering at the charge ordering wavevector.
In addition, related results can be obtained in the case of spin ordering, where again, depending on
the pattern (spins on the copper sites, or spins on the oxygen sites), one sees a significant difference
in the dichroism.  The latter is particularly relevant, since resonant
x-ray scattering at the magnetic wavevector of underdoped cuprates is not possible at the Cu L$_2$ 
and L$_3$ edges because of the kinematic constraint, making absorption the only ideal x-ray
probe of the spin order \cite{spin}.

In summary, linear dichroism is an exquisite probe of charge order in cuprates, and can be used to
determine the actual modulation pattern due to the site sensitivity of resonant x-rays.  Similar considerations
also apply to the incommensurate spin order seen for even more underdoped samples.

The author thanks Yves Joly and Sergio Di Matteo for several helpful discussions.
This work was supported by the Materials Sciences and Engineering
Division, Basic Energy Sciences, Office of Science, US DOE.


\begin{thebibliography}{99}
\bibitem{shift}
The calculated peaks are shifted in energy relative to experiment, though the energy separation of the L$_3$ and L$_2$
peaks is the same.
\bibitem{spin}
Resonant x-ray scattering studies of the spin order can be done at the Cu K edge, but the effects are weak since 
only the orbital moment contribution on the copper sites can be probed.
\end{thebibliography}
\end{document}